\documentstyle[prl,aps,twocolumn]{revtex}

\begin{document}
\draft
\title{THEORY\ OF\ RANDOM\ MATRICES\ WITH\ STRONG\ LEVEL\ CONFINEMENT}
\author{{\bf V. Freilikher}$\dagger $, {\bf E. Kanzieper}$\dagger $,{\bf \ and I.
Yurkevich}$\ddagger $}
\address{$\dagger $The Jack and Pearl Resnick Institute of Advanced Technology,\\
Department of Physics, Bar-Ilan University, Ramat-Gan 52900, Israel\\
$\ddagger $International Centre for Theoretical Physics, Trieste 34100, Italy}
\date{September 11, 1995}
\maketitle

\begin{abstract}
Unitary ensembles of large $n\times n$ random matrices with a non-Gaussian
probability distribution $P\left[ {\bf H}\right] \propto \exp \left\{ -\text{%
tr}V\left[ {\bf H}\right] \right\} $ are studied using a theory of
polynomials orthogonal with respect to exponential weights. Asymptotically
exact expressions for density of levels, one- and two-point Green's
functions are calculated. We show that in the large-$n$ limit the properly
rescaled local eigenvalue correlations are independent of $P\left[ {\bf H}%
\right] $ while global smoothed connected correlations depend on $P\left[ 
{\bf H}\right] $ only through the endpoints of spectrum. We also establish
previously unknown intimate connection between structure of Szeg\"o function
entering strong polynomial asymptotics and mean-field equation by Dyson.
\end{abstract}

\pacs{{\tt cond-mat/9510002}}

\narrowtext

Random-matrix theory turned out to be general phenomenological approach to
description of various phenomena in complex physical systems \cite{Mehta}
such as complex nuclei, mesoscopic conductors or quantum chaotic systems.
Due to complicated nature of these objects their treatment, based on the
direct integration of microscopic equations, runs across considerable
obstacles. In this situation the more reasonable way is to deal with the
joint distribution function $P\left[ {\bf H}\right] $ of the matrix elements
of Hamiltonian ${\bf H}$ describing the system involved. If there is not
preferential basis in the space of matrix elements ${\bf H}_{ij}$ (i.e. the
system in question is ``as random as possible'') one has to require $P\left[ 
{\bf H}\right] d\left[ {\bf H}\right] $ to be invariant under ${\cal R}$%
-transformation of the Hamiltonian: ${\bf H}\rightarrow {\cal R}^{-1}{\bf H}%
{\cal R}$. Here ${\cal R}$ is orthogonal, unitary or symplectic matrix
reflecting the fundamental symmetry of the underlying Hamiltonian. The
general form of $P\left[ {\bf H}\right] $ compatible with invariance
requirement can be written as $P\left[ {\bf H}\right] =Z^{-1}\exp \left\{ -%
\text{tr}V\left[ {\bf H}\right] \right\} $ with $V\left[ {\bf H}\right] $
providing existence of partition function $Z$. Transformation of ${\bf H}$
to diagonal form ${\bf H}={\cal S}_\beta ^{-1}{\bf X}{\cal S}_\beta $ and
further integration over orthogonal $\left( \beta =1\right) $, unitary $%
\left( \beta =2\right) $ or symplectic $\left( \beta =4\right) $ group $d\mu
\left( {\cal S}_\beta \right) $ in the construction $P\left[ {\bf H}\right]
d\left[ {\bf H}\right] $ yield famous expression for the joint probability
density function of the eigenvalues $\left\{ x\right\} $ of the matrix ${\bf %
H}$: 
\begin{equation}
P\left( \left\{ x\right\} \right) =\frac 1{Z_{n\beta }}\exp \left\{ -\beta
\left[ \sum_iV\left( x_i\right) -\sum_{i<j}\ln \left| x_i-x_j\right| \right]
\right\} \text{.}  \label{i.01}
\end{equation}
Confinement potential $V\left( x\right) $ contains, in an integral way, an
information about correlations between different matrix elements of a random
Hamiltonian ${\bf H}$ while logarithmic interaction induced by the
invariance requirement is usually treated as a pairwise repulsion between
energy levels. We note that in Eq. (\ref{i.01}) the symmetry parameter $%
\beta $ is factored out from $V\left( x\right) $ to fix density of levels in
random-matrix ensembles with the same confinement potential but with
different underlying symmetry.

The structure of Eq. (\ref{i.01}) allows to represent exactly all the global
and local statistical characteristics of the physical system with broken
time-reversal symmetry $\left( \beta =2\right) $ in the terms of polynomials
orthogonal with respect to the measure $d\alpha \left( x\right) =\exp \left(
-2V\left( x\right) \right) dx$.

Hitherto only few cases of strong confinement potentials were {\it rigorously%
} treated in the random-matrix theory. Random-matrix ensembles associated
with classical orthogonal polynomials have been considered in \cite{Nagao};
those related to some non-classical orthogonal polynomials have been treated
in \cite{Pastur}, \cite{FKY}. We also mention the basic work \cite
{Brezin-Zee} where {\it ansatz} for functional form of polynomial
asymptotics was proposed to describe a class of non-Gaussian unitary
random-matrix ensembles.

Since generally an analytical calculation of the corresponding set of
orthogonal polynomials is a non-trivial problem, a number of alternative
approximate methods were developed. The mean-field approximation proposed by
Dyson \cite{Dyson} allows to calculate density of levels in random-matrix
ensemble. This approach being combined with the method of functional
derivative of Beenakker \cite{Beenakker} makes it possible to compute global
(smoothed) eigenvalue correlations in large random matrices. Smoothed
correlations can also be obtained by diagrammatic approach of Br\'ezin and
Zee \cite{Brezin-Zee-diagrams} or by invoking the arguments based on
macroscopic electrostatics and linear response theory \cite{Forrester}. We
stress that all the methods mentioned above allow to study correlations only
in {\it long-range regime} and, in this sense, they are less informative as
compared with the method of orthogonal polynomials \cite{Brezin-Zee}. It is
worth pointing out the supersymmetry formalism \cite{Weidenmuller}, recently
developed for invariant random-matrix ensembles with non-Gaussian
probability distribution function $P\left[ {\bf H}\right] $, which is
exceptional in that it allows to investigate {\it local} eigenvalue
correlations and represents a powerful alternative approach to the classical
method of orthogonal polynomials.

It is the aim of this Letter to outline how the problem of non-Gaussian
ensembles with unitary symmetry can be handled by the method of orthogonal
polynomials in rather general case. Extended version of the article
containing detailed mathematical calculations will be published elsewhere 
\cite{To be}. We stress that our treatment is exact (i.e. it does not
involve any conjectures) and based on the recent results \cite{AAM-1993}
obtained in the theory of polynomials orthogonal with respect to exponential
weights on {\bf R}.

Let us consider a class of even confinement potentials $V\left( x\right) $
supported on the whole real axis $x\in \left( -\infty ,+\infty \right) $
which are {\it of smooth polynomial growth at infinity and increase there at
least as }$\left| x\right| ^{1+\delta }$ ($\delta $ is arbitrary small
positive number). More precisely, we demand $V\left( x\right) $ and $%
d^2V/dx^2$ be continuous in $x\in \left( 0,+\infty \right) $, and $dV/dx>0$
in the same domain of variable $x$. Further we define the function $T\left(
x\right) =1+x\cdot d^2V/dx^2\cdot \left( dV/dx\right) ^{-1}$ and assume that
for some $A>1$ and $B>1$ the inequality $A\leq T\left( x\right) \leq B$
holds for $x\in \left( 0,+\infty \right) $, and also for $x$ positive and
large enough $x^2\cdot \left| d^3V/dx^3\right| \cdot \left( dV/dx\right)
^{-1}\leq const$. The class of potentials $V\left( x\right) $ satisfying all
the above requirements is said to be of the {\it Freud type}. The typical
examples of the Freud potentials are (i) $V\left( x\right) =\left| x\right|
^\alpha $ with $\alpha >1$; (ii) $V\left( x\right) =\left| x\right| ^\alpha
\ln ^\beta \left( \gamma +x^2\right) $ with $\alpha >1,$ $\beta \in {\bf R}$%
, and $\gamma $ large enough.

Defining the Freud (non-Gaussian) measure $d\alpha _{{\cal F}}\left(
x\right) =$ $w_{{\cal F}}\left( x\right) dx=\exp \left( -2V\left( x\right)
\right) dx$ we can introduce a set of orthogonal polynomials $P_n\left(
x\right) $ 
\begin{equation}
\int_{-\infty }^{+\infty }P_n\left( x\right) P_m\left( x\right) d\alpha _{%
{\cal F}}\left( x\right) =\delta _{nm}\text{,}  \label{eq.03}
\end{equation}
for which the following basic result was obtained by D. S. Lubinsky \cite
{AAM-1993}: 
\[
\lim_{n\rightarrow \infty }\int_{-1}^{+1}d\lambda \left\{ a_n^{1/2}P_n\left(
a_n\lambda \right) \right. 
\]
\begin{equation}
\left. -\left( \frac 2\pi \right) ^{1/2}Re\left[ z^nD^{-2}\left( F_n;\frac 1z%
\right) \right] \right\} ^2w_{{\cal F}}\left( a_n\lambda \right) =0\text{.}
\label{eq.04}
\end{equation}
Here parametrizations $z=e^{i\theta }$ and $\lambda =\cos \theta $ were used.

Szeg\"o function $D\left( g;z\right) $, appeared in Eq. (\ref{eq.04}), is of
fundamental importance in the whole theory of orthogonal polynomials \cite
{Szego-1921}, and takes the form 
\begin{equation}
D\left( g;z\right) =\exp \left( \frac 1{4\pi }\int_{-\pi }^{+\pi }d\varphi 
\frac{1+ze^{-i\varphi }}{1-ze^{-i\varphi }}\ln g\left( \varphi \right)
\right) \text{.}  \label{eq.06}
\end{equation}
The first argument of Szeg\"o function in Eq. (\ref{eq.04}) is 
\begin{equation}
F_n\left( \varphi \right) =\exp \left( -V\left( a_n\cos \varphi \right)
\right) \left| \sin \varphi \right| ^{1/2}\text{,}  \label{eq.07}
\end{equation}
and $a_n$ is the $n$-th Mhaskar-Rahmanov-Saff number being the positive root
of the integral equation \cite{MRS-numb} 
\begin{equation}
n=\frac{2a_n}\pi \int_0^1\frac{\lambda d\lambda }{\sqrt{1-\lambda ^2}}\left( 
\frac{dV}{dx}\right) _{x=a_n\lambda }\text{.}  \label{eq.05}
\end{equation}
(In what follows it will be seen that $a_n$ is none other than band edge for
eigenvalues of corresponding random-matrix ensemble).

Passing on to the integration over $x=a_n\lambda $ in Eqs. (\ref{eq.04}) and
(\ref{eq.05}), and leaving aside the problem of remainder term we arrive at
the asymptotic formula, $n\gg 1$, for orthogonal polynomials of the Freud
type:

\begin{equation}
P_n\left( x\right) =\sqrt{\frac 2{\pi a_n}}Re\left[ z^nD^{-2}\left( F_n;%
\frac 1z\right) \right] ,\text{ }x\in \left( -a_n,+a_n\right) \text{.}
\label{eq.10}
\end{equation}
Making use of the properties of Szeg\"o function $D\left( g;e^{i\theta
}\right) $ \cite{Szego-monograph} one can obtain the following formula for
the orthonormal ``wave functions'' $\psi _n\left( x\right) =P_n\left(
x\right) \exp (-V\left( x\right) )$ within the interval $x\in \left(
-a_n,+a_n\right) $ in the large-$n$ limit: 
\begin{equation}
\psi _n\left( x\right) =\sqrt{\frac 2{\pi a_n}}\left[ 1-\left( \frac x{a_n}%
\right) ^2\right] ^{-1/4}\cos \Phi _n\left( x\right) \text{,}  \label{eq.22}
\end{equation}
where 
\begin{equation}
\Phi _n\left( x\right) =\gamma _n\left( x\right) +n\arccos \left( \frac x{a_n%
}\right) \text{,}  \label{eq.300}
\end{equation}
\begin{equation}
\gamma _n\left( x\right) =\frac 1{2\pi }{\cal P}\int_{-a_n}^{+a_n}d\xi \frac{%
\sqrt{a_n^2-x^2}}{\sqrt{a_n^2-\xi ^2}}\frac{h\left( \xi \right) }{\xi -x}%
\text{,}  \label{eq.301}
\end{equation}
\begin{equation}
h\left( \xi \right) =\frac 12\ln \left[ 1-\left( \frac \xi {a_n}\right)
^2\right] -2V\left( \xi \right) \text{.}  \label{eq.302}
\end{equation}
In Eq. (\ref{eq.301}) ${\cal P}$ stands for principal value of the integral.
Since the same formula takes place for the {\it Erd\"os-type }confinement
potentials \cite{Erdos} which {\it grow faster than any polynomial at
infinity} (see \cite{PRN}, Ch. 2) the further treatment applies equally to
unitary random-matrix ensembles with confinement potentials of the Freud-
and Erd\"os-type. The simple examples of Erd\"os-type confinement potentials
are (i) $V\left( x\right) =\exp _k\left( \left| x\right| ^\alpha \right) $
with $\alpha >0$ and $k\geq 1$ (here $\exp _k$ denotes the exponent iterated 
$k$-times); (ii) $V\left( x\right) =\exp \left( \ln ^\alpha \left( \gamma
+x^2\right) \right) $ with $\alpha >1$, and $\gamma $ large enough.

All the global and local characteristics of the random-matrix ensembles are
determined by two-point kernel \cite{Mehta}

\begin{equation}
K_n\left( x,y\right) =\frac{k_{n-1}}{k_n}\frac{\psi _n\left( y\right) \psi
_{n-1}\left( x\right) -\psi _n\left( x\right) \psi _{n-1}\left( y\right) }{%
y-x}\text{,}  \label{eq.23}
\end{equation}
where $k_n$ is a leading coefficient of the orthogonal polynomial $P_n\left(
x\right) $. Substituting Eq. (\ref{eq.22}) into Eq. (\ref{eq.23}) we obtain
after straightforward calculations in the large-$n$ limit:

\[
K_n\left( x,y\right) =\frac 1{\pi \left( y-x\right) }\left[ f_n\left(
x\right) f_n\left( y\right) \right] ^{-1/4} 
\]
\[
\times \left\{ \cos \Phi _n\left( x\right) \cos \Phi _n\left( y\right) \frac{%
x-y}{a_n}\right. 
\]
\[
-\sin \Phi _n\left( y\right) \cos \Phi _n\left( x\right) \sqrt{f_n\left(
y\right) } 
\]
\begin{equation}
\left. +\sin \Phi _n\left( x\right) \cos \Phi _n\left( y\right) \sqrt{%
f_n\left( x\right) }\right\} \text{,}  \label{eq.28}
\end{equation}
where $f_n\left( x\right) =1-\left( x/a_n\right) ^2$. This equation is valid
for arbitrary $x$ and $y$ lying within the band $\left( -a_n,a_n\right) $.

Equation (\ref{eq.28}) allows to compute smoothed (over the rapid
oscillations) connected correlations $\nu _c\left( x,y\right) $ \cite
{Brezin-Zee} $\nu _c\left( x,y\right) =\overline{\left\langle \nu _n\left(
x\right) \nu _n\left( y\right) \right\rangle }-\overline{\left\langle \nu
_n\left( x\right) \right\rangle \left\langle \nu _n\left( y\right)
\right\rangle }=-\overline{K_n^2\left( x,y\right) }$, where $x\neq y$ and $%
\overline{\left( ...\right) }$ denotes averaging over intervals $\left|
\Delta x\right| \ll a_n$ and $\left| \Delta y\right| \ll a_n$ but still
containing many eigenlevels. Direct calculations yield the simple universal
relationship 
\begin{equation}
\nu _c\left( x,y\right) =-\frac 1{2\pi ^2}\frac{a_n^2-xy}{\left( x-y\right)
^2\sqrt{a_n^2-x^2}\sqrt{a_n^2-y^2}},\text{ }x\neq y  \label{eq.30}
\end{equation}
with dependence on the potential $V\left( x\right) $ only through the
endpoint $a_n$ of the spectrum.

Now we turn to the local properties of two-point kernel. Assuming that in
Eq. (\ref{eq.28}) $\left| x-y\right| \ll a_n$, and both $x$ and $y$ stay
away from the band edge $a_n$ we obtain 
\begin{equation}
K_n\left( x,y\right) =\frac{\sin \left( \Phi _n\left( x\right) -\Phi
_n\left( y\right) \right) }{\pi \left( y-x\right) }\text{.}  \label{eq.35}
\end{equation}
This two-point kernel may be rewritten in locally universal form noting that
following representation for $\Phi _n\left( x\right) $ takes place 
\begin{equation}
\Phi _n\left( x\right) =\frac 12\arccos \left( \frac x{a_n}\right) -\pi
\int_0^x\omega _{a_n}\left( \xi \right) d\xi +\frac \pi 4\left( 2n-1\right) 
\text{,}  \label{eq.45}
\end{equation}
where 
\begin{equation}
\omega _{a_n}\left( x\right) =\frac 2{\pi ^2}{\cal P}\int_0^{a_n}\frac{\xi
d\xi }{\xi ^2-x^2}\frac{dV}{d\xi }\frac{\sqrt{a_n^2-x^2}}{\sqrt{a_n^2-\xi ^2}%
}\text{.}  \label{eq.42}
\end{equation}
Since the characteristic scale of the changing of $\omega _{a_n}\left( \xi
\right) $ is $\varsigma \sim \left( \omega _{a_n}^{-1}d\omega _{a_n}/d\xi
\right) ^{-1}\sim a_n$, Eq. (\ref{eq.35}) for $\left| x-y\right| \ll
\varsigma $ is reduced to universal form 
\begin{equation}
K_n\left( x,y\right) =\frac{\sin \left[ \pi \overline{\nu }_n\left(
y-x\right) \right] }{\pi \left( y-x\right) }  \label{eq.47}
\end{equation}
with $\overline{\nu }_n=\omega _{a_n}\left( \frac{x+y}2\right) $ playing the
role of local density of levels. Correspondingly, local two-level cluster
function being rewritten in rescaled variables $s$ and $s^{\prime }$ 
\begin{equation}
Y_2\left( s,s^{\prime }\right) =\left( \frac{K_n^2\left( x,y\right) }{%
\left\langle \nu _n\left( x\right) \right\rangle \left\langle \nu _n\left(
y\right) \right\rangle }\right) _{%
{x=x\left( s\right)  \atopwithdelims.. y=y\left( s^{\prime }\right) }
}=\frac{\sin ^2\left[ \pi \left( s-s^{\prime }\right) \right] }{\left[ \pi
\left( s-s^{\prime }\right) \right] ^2}  \label{eq.48}
\end{equation}
proves universal Wigner-Dyson level statistics in the unitary random-matrix
ensembles with Freud- and Erd\"os-type confinement potentials (here $s=%
\overline{\nu }_nx$ and $s^{\prime }=\overline{\nu }_ny$ are the eigenvalues
measured in the local mean level-spacing).

Expression for density of levels $\left\langle \nu _n\left( x\right)
\right\rangle =K_n\left( x,x\right) $ immediately follows from Eq. (\ref
{eq.35}), $\left\langle \nu _n\left( x\right) \right\rangle =-\frac 1\pi
d\Phi _n/dx$. It can be written down in two equivalent forms. The first one, 
\[
\left\langle \nu _n\left( x=a_n\cos \theta \right) \right\rangle =\frac 1{%
\pi a_n\sin \theta } 
\]
\begin{equation}
\times \frac d{d\theta }\left[ n\theta +\arg D\left( e^{-2V\left( a_n\cos
\varphi \right) }\left| \sin \varphi \right| ;e^{i\theta }\right) \right] 
\text{,}  \label{eq.51}
\end{equation}
establishes connection between density of levels in random-matrix ensembles
with confinement potentials of Freud- and Erd\"os-type, and Szeg\"o
function, Eq. (\ref{eq.06}), for corresponding set of orthogonal polynomials.

Another representation of the density of states reads 
\begin{equation}
\left\langle \nu _n\left( x\right) \right\rangle =\frac 2{\pi ^2}{\cal P}%
\int_0^{a_n}\frac{\xi d\xi }{\xi ^2-x^2}\frac{dV}{d\xi }\frac{\sqrt{a_n^2-x^2%
}}{\sqrt{a_n^2-\xi ^2}}\text{.}  \label{eq.52}
\end{equation}
Considering this expression as an equation for $V\left( x\right) $ one can
resolve it invoking the theory of integral equations with Cauchy kernel \cite
{Akhiezer}: 
\begin{equation}
V\left( x\right) =\int_{-a_n}^{+a_n}dx^{\prime }\left\langle \nu _n\left(
x^{\prime }\right) \right\rangle \ln \left| x-x^{\prime }\right| +\mu
\label{eq.52b}
\end{equation}
with $\mu $ being ``chemical potential''. It is non more than famous
mean-field equation which, in our treatment, finally follows from the
asymptotic formula Eq. (\ref{eq.10}) for orthogonal polynomials without any
physical speculations. Thus, quite surprisingly, Szeg\"o function, Eq. (\ref
{eq.06}), turns out to be closely related to the mean-field approximation by
Dyson \cite{Dyson}.

Now the calculation of the one-point Green's function 
\[
G^p\left( x\right) =\left\langle \text{tr}\frac 1{x-{\bf H}+ip0}%
\right\rangle 
\]
\begin{equation}
={\cal P}\int_{-a_n}^{+a_n}d\xi \frac{\left\langle \nu _n\left( \xi \right)
\right\rangle }{x-\xi }-i\pi p\left\langle \nu _n\left( x\right)
\right\rangle  \label{eq.53}
\end{equation}
with $p=\pm 1$ becomes quite trivial. We obtain by means of Eqs. (\ref{eq.52}%
) and (\ref{eq.52b}): 
\begin{equation}
G^p\left( x\right) =\frac{dV}{dx}-\frac{2ip}\pi {\cal P}\int_0^{a_n}\frac{%
\xi d\xi }{\xi ^2-x^2}\frac{dV}{d\xi }\frac{\sqrt{a_n^2-x^2}}{\sqrt{%
a_n^2-\xi ^2}}\text{.}  \label{eq.55}
\end{equation}

Two-point connected Green's function 
\[
G_c^{pp^{\prime }}\left( x,x^{\prime }\right) =\left\langle \text{tr}\frac 1{%
x_p-{\bf H}}\text{tr}\frac 1{x_{p^{\prime }}^{\prime }-{\bf H}}\right\rangle 
\]
\begin{equation}
-\left\langle \text{tr}\frac 1{x_p-{\bf H}}\right\rangle \left\langle \text{%
tr}\frac 1{x_{p^{\prime }}^{\prime }-{\bf H}}\right\rangle \text{,}
\label{eq.200}
\end{equation}
where $x_p=x+ip0$ and $x_{p^{\prime }}^{\prime }=x^{\prime }+ip^{\prime }0$ (%
$p,p^{\prime }=\pm 1$), can also be computed within the proposed formalism.
These rather cumbersome calculations \cite{To be} lead to the following
formula for the smoothed connected two-point Green's function: 
\[
\overline{G_c^{pp^{\prime }}\left( x,x^{\prime }\right) }=\frac 12\left\{
pp^{\prime }\frac{a_n^2-xx^{\prime }}{\left( x-x^{\prime }\right) ^2\sqrt{%
a_n^2-x^2}\sqrt{a_n^2-x^{\prime 2}}}\right. 
\]
\begin{equation}
\left. -\frac 1{\left( x_p-x_{p^{\prime }}^{\prime }\right) ^2}\right\} ,%
\text{ }x\neq x^{\prime }.  \label{eq.213}
\end{equation}

It also can be shown that the two-point Green's function in the
random-matrix theory exhibits the local universality. Introducing normalized
and rescaled two-point connected Green's function $g_c^{pp^{\prime }}\left(
s,s^{\prime }\right) =G_c^{pp^{\prime }}\left( x,x^{\prime }\right)
/\left\langle \nu _n\left( x\right) \right\rangle \left\langle \nu _n\left(
x^{\prime }\right) \right\rangle $ expressed in the terms of the eigenvalues 
$s$ and $s^{\prime }$ measured in the local mean level-spacing, we obtain
the following universal relationship that does not depend on the
distribution function $P\left[ {\bf H}\right] $: 
\[
g_c^{pp^{\prime }}\left( s,s^{\prime }\right) =\pi ^2\left| p-p^{\prime
}\right| \delta \left( s-s^{\prime }\right) 
\]
\begin{equation}
+i\left( p-p^{\prime }\right) \frac{\sin \left[ \pi \left( s-s^{\prime
}\right) \right] }{\left( s-s^{\prime }\right) ^2}\text{e}^{i\pi \left(
s-s^{\prime }\right) \text{sign}\left( p-p^{\prime }\right) }\text{.}
\label{eq.217}
\end{equation}
Note that expression of this type was previously obtained in \cite{Zuk} only
for Gaussian random-matrix ensemble using supersymmetry formalism.

In summary, we have presented rigorous analytical consideration of the basic
problems of the theory of large Hermitian random matrices with non-Gaussian
distribution function $P\left[ {\bf H}\right] \propto \exp \left\{ -\text{tr}%
V\left[ {\bf H}\right] \right\} $. Our treatment is applied to very wide
class of monotonous confinement potentials which behave {\it at least as }$%
\left| x\right| ^{1+\delta }${\it \ }$\left( \delta >0\right) ${\it \ and
can grow as or even faster than any polynomial at infinity} (note, that this
class of confinement potentials is much richer than that considered in \cite
{Brezin-Zee}). We have shown that in such unitary random-matrix ensembles
with non-Gaussian measure $P\left[ {\bf H}\right] d\left[ {\bf H}\right] $
the density of levels and one-point Green's function essentially depend on
the measure. In contrast, (connected) two-point characteristics of spectrum
are rather universal. Indeed, we have observed global universality of
smoothed two-point connected correlators and local universality of those
without smoothing over rapid oscillations. The correlators were shown to
depend on the measure only through the endpoints of spectrum (global
universality) or through the local density of levels (local universality).
Presented rigorous polynomial analysis enabled us to establish new local
universal relationship in the random-matrix theory for normalized and
rescaled connected two-point Green's function $g_c^{pp^{\prime }}\left(
s,s^{\prime }\right) $. It is also worthy of notice an interesting and quite
surprising connection between the structure of Szeg\"o function and
mean-field equation that has been revealed in the proposed formalism. In our
opinion this fact is far from obvious.

One of the authors (E. K.) gratefully acknowledges financial support of The
Ministry of Science and The Arts of Israel.


\begin{references}
\bibitem{Erdos}  Namely, $V\left( x\right) $ should be even and continuous
in $x\in \left( -\infty ,+\infty \right) $, $d^2V/dx^2$ be continuous in $%
x\in \left( 0,+\infty \right) $, $dV/dx>0$ in the same domain of $x$ and
continuous at $x=0$, while function $T\left( x\right) $ be positive and
increasing in $x\in \left( 0,+\infty \right) $ with $\lim_{x\rightarrow
+0}T\left( x\right) >0$, $\lim_{x\rightarrow \infty }T\left( x\right)
=\infty $, and $T\left( x\right) ={\cal O}\left( \left( dV/dx\right)
^{1/15}\right) $ as $x\rightarrow \infty $. The requirements $\left(
d^2V/dx^2\right) \cdot \left( dV/dx\right) ^{-1}\sim V\left( x\right)
^{-1}\cdot \left( dV/dx\right) $ and $\left| d^3V/dx^3\right| \cdot \left(
dV/dx\right) ^{-1}\leq const\cdot \left( V\left( x\right) ^{-1}\cdot \left(
dV/dx\right) \right) ^2$ must be fulfilled for $x$ large enough as well.
\end{references}
\end{document}